\documentclass[12pt,a4paper]{amsart}
\usepackage{amscd,amssymb,amsthm}
\usepackage{jipam}
\usepackage{url}
\newcommand{\bO}{{\mathcal O}}
\newcommand{\GO}{{\mathcal O}}
\def\C{\mathbb C}
\newcommand{\Z}{\ensuremath{\mathbb Z}}

\usepackage{graphicx,url}
\usepackage{enumerate}
\usepackage[ruled,section]{algorithm}
\usepackage{algorithmic}

\title[Characteristic and minimal polynomials coefficient bounds]{Bounds on the coefficients of the characteristic and minimal polynomials}
\author{Jean-Guillaume Dumas}
\date{\today}
\address{Laboratoire Jean Kuntzmann, Universit\'e
  Joseph Fourier, Grenoble I}
\address{UMR CNRS 5224, 51 avenue des Math\'ematiques, 38041 Grenoble, France.}
\email{Jean-Guillaume.Dumas@imag.fr}
\urladdr{http://ljk.imag.fr/membres/Jean-Guillaume.Dumas}

\DeclareMathOperator{\minpoly}{\textup{minpoly}}
\DeclareMathOperator{\cpcb}{0.21163175}

\begin{document}
\begin{abstract}
This note presents absolute bounds on the size of the coefficients of the
characteristic and minimal polynomials depending on the size of the
coefficients of the associated matrix. Moreover, we present algorithms to
compute more precise input-dependant bounds on these coefficients.
Such bounds are e.g. useful to perform deterministic Chinese
remaindering of the characteristic or minimal polynomial of an integer
matrix.
\end{abstract}
\maketitle 

\section{Introduction}
The Frobenius normal form of a matrix is used to test
two matrices for similarity. Although the Frobenius normal form contains more in
formation on 
the matrix than the characteristic polynomial, most efficient algorithms to compute it are
 based on  computations of characteristic polynomial (see for example
\cite[\S 9.7]{Storjohann:2000:thesis}).
Now the Smith normal form of an integer matrix is useful e.g. in the
computation of homology groups and its computation can be done via the
integer minimal polynomial \cite{jgd:2001:JSC}. 

In both cases, the polynomials are computed first modulo several prime
numbers and then only reconstructed via Chinese
remaindering \cite[Theorem 10.25]{VonzurGathen:1999:MCA}.
Thus, precise bounds on the integer coefficients of the integer
characteristic or minimal polynomials of an integer matrix are used to
know how many primes are sufficient to perform a Chinese remaindering
of the modularly computed polynomials. Some bounds on the minimal
polynomial coefficients, respectively the characteristic polynomial,
have been presented in \cite{jgd:2001:JSC}, respectively in
\cite{jgd:2005:charp}. The aim of this note is to present sharper
estimates in both cases. 

For both polynomials we present two kind of
results:
{\em absolute estimates}, useful to compare complexity constants, and
{\em algorithms} which compute more precise estimates based on the
properties of the input matrix discovered at runtime. Of course, the
goal is to provide such estimates at a cost
negligible when compared to that of actually computing the
polynomials.
\section{Bound on the minors for the characteristic polynomial}
\subsection{Hadamard's bound on the minors}
The first bound of the characteristic polynomial coefficient 
uses Hadamard's bound, $|det(A)| \leq \sqrt{nB^2}^n$, see e.g. \cite[Theorem 16.6]{VonzurGathen:1999:MCA},
to show that the coefficients of the characteristic polynomial
could be larger, but only slightly:
\begin{lemma}\label{lem:hadamard}
Let $A \in \C^{n \times n}$, with $n \geq 4$, whose coefficients are bounded in absolute
value by $B>1$. The coefficients of the characteristic polynomial $C_A$
of $A$ are denoted by $c_j$, $j=0..n$ and $||C_A||_\infty = max \{
|c_j| \}$. Then
$$\log_2( ||C_A||_\infty) \leq \frac{n}{2}\left(\log_2(n)+\log_2(B^2)+\cpcb\right)$$
\end{lemma}
\begin{proof}
Observe that $c_j$, the $j$-th coefficient of the characteristic polynomial, is an alternate
sum of all the $(n-j) \times (n-j)$ diagonal minors of $A$, see
e.g. \cite[\S III.7]{Gantmacher:1959:TMone}. It is therefore bounded
by $F(n,j)={n \choose j} \sqrt{(n-j)B^2}^{(n-j)}$. First note, that from the symmetry of the
binomial coefficients we only need to explore the $\lfloor n/2
\rfloor$ first ones, since 
$\sqrt{(n-j)B^2}^{(n-j)} > \sqrt{j B^2}^{j}$ for 
$j<\lfloor n/2 \rfloor$. 

The lemma is true for $j=0$ by Hadamard's bound.

For $j=1$ and $n\geq 2$, we set 
$f(n) = \frac{2}{n}\left( \log_2 \left( F(n,1) \right) - \frac{n}{2}
  \log_2(n)-(n-1)\log_2(B) \right)$.
Now $\frac{df}{dn} = \frac{2n-2+n\ln(n-1)-2n\ln(n)+2\ln(n)-ln(n-1)}{n^2(n-1)\ln(2)}$.
Thus, the numerator of the derivative of $f(n)$ has two roots, one below $2$ and one between
$6$ and $7$.
Also, $f(n)$ is increasing from $2$ to the second root and
decreasing afterwards. With $n \geq 4$, the maximal value of $f(n)$
is therefore at $n=6$, for which it is
$\frac{5}{6}\log_2(5) -\frac{2}{3}\log_2(6)< \cpcb$.

For other $j$'s, Stirling's formula has
been extended for the binomial coefficient by St\u{a}nic\u{a} in
\cite{Stanica:2001:binomial}, and gives $\forall i\geq 2$,
\[ 
{n \choose j} <
\frac{e^{\frac{1}{12n}-\frac{1}{12j+1}-\frac{1}{12(n-j)+1}}}{\sqrt{2\pi}}\sqrt{\frac{n}{j(n-j)}}\left(\frac{n}{j}\right)^j\left(\frac{n}{n-j}\right)^{n-j}.
\]

Now first $\frac{1}{12n}-\frac{1}{12j+1}-\frac{1}{12(n-j)+1} <
\frac{1}{12n}-\frac{2}{6n+1}$, since the maximal value of the latter
is at $j=\frac{n}{2}$.
Therefore, $\log_2
\left(\frac{e^{\frac{1}{12n}-\frac{1}{12j+1}-\frac{1}{12(n-j)+1}}}{\sqrt{2\pi}}\right)
\leq \log_2
\left(\frac{1}{\sqrt{2\pi}}\right) < -1.325$.

Then
$\frac{n}{j(n-j)}$ is decreasing in $j$ for $2 \leq j<\lfloor n/2\rfloor$ so that its
maximum is $\frac{n}{2(n-2)}$. 
\\
Consider now the rest of the approximation
$K(n,j) = \left(\frac{n}{j}\right)^j\left(\frac{n}{n-j}\right)^{n-j}\sqrt{(n-j)B^2}^{(n-j)}.$
We have $\log_2( K(n,j) ) = \frac{n-j}{2}\log_2(B^2) +
\frac{n}{2}\log_2(n) + \frac{n}{2}T(n,j)$, where 
$T(n,j) = \log_2(\frac{n}{n-j})+\frac{j}{n}\log_2(\frac{n-j}{j^2})$.
Well $T(n,j)$ is maximal for $j=\frac{-1+\sqrt{1+4en}}{2e}$. 
We end with the fact that for $n \geq 4$, 
$T\left(n, \frac{-1+\sqrt{1+4en}}{2e}\right) -\frac{2}{n}\log_2(\sqrt{2\pi})+\frac{1}{n}\log_2\left(\frac{n}{2(n-2)}\right)$ is
maximal over $\Z$ for $n=16$ where it is lower than
$0.2052$. The latter is lower than $\cpcb$.
\end{proof}
We show the effectiveness of our bound on an example matrix:
\begin{equation}\label{matex}\left[\begin{smallmatrix}
1&1&1&1&1\\
1&1&-1&-1&-1\\
1&-1&1&-1&-1\\
1&-1&-1&1&-1\\
1&-1&-1&-1&1
\end{smallmatrix}\right].\end{equation}
This matrix has {{$X^5-5X^4+40X^2-80X+48$}}  for characteristic polynomial and $80 = {5 \choose 1} \sqrt{4}^4$ is
greater than Hadamard's bound $55.9$, and less than our bound $80.66661$.

Note that this numerical bound improves on the one used in
\cite[lemma 2.1]{Giesbrecht:2002:CRF} since $\cpcb <
2+\text{log}_2(e)\approx 3.4427$. While yielding the same asymptotic
result, their bound would state e.g. that the
coefficients of the characteristic polynomial of the example 
are lower than $21793$.

\subsection{Locating the largest coefficient}
The proof of lemma \ref{lem:hadamard} suggests that the largest
coefficient is to be found between the $\GO(\sqrt{n})$ last ones. 
In next lemma we take $B$ into account in order to sharpen this localization.
This gives a simple search procedure computing a more accurate bound 
on the fly, as
soon as $B$ is known.
\begin{lemma}\label{lem:comput}
 Let $A \in \C^{n \times n}$, with $n \geq 4$, whose coefficients are bounded in absolute
value by $B>1$. The characteristic polynomial
of $A$ is $C_A$. Then
$$||C_A||_\infty \leq max_{i=0..\frac{-1+\sqrt{1+2\delta B^2n}}{\delta
    B^2}} {n \choose i}
\sqrt{(n-i)B^2}^{(n-i)}$$ where $\delta \approx 5.418236$.
Moreover, the cost of computing the associated bound on the size is
$$\GO\left(\frac{\sqrt{n}}{B}\right).$$
\end{lemma}
This localization improves by a
factor close to $\frac{1}{B}$, the localization of the largest coefficient
proposed in \cite[Lemma 4.1]{jgd:2005:charp}.
\begin{proof}
Consider 
$F(n,j)={n \choose j} \sqrt{(n-j)B^2}^{(n-j)}$ for
$j=2..\lfloor \frac{n}{2} \rfloor$.
The numerator of the derivative of $F$ with respect to $j$ 
is
$$n!\sqrt{(n-j)B^2}^{n-j}\left(2H(n-j)-2H(j)-\ln(n-j)-\ln(B^2)-1\right)$$
where $H(k)=\sum_{l=1}^{k} \frac{1}{l}$ is the $k$-th Harmonic number.
We have the bounds 
$\ln(k)+\gamma +\frac{1}{2k+\frac{1}{1-\gamma}-2}< H(k) < \ln(k)+\gamma +\frac{1}{2k+\frac{1}{3}}$ from
\cite[Theorem 2]{Qi:2005:harmonic}.
This bounds proves that $F(n,j)$ has at most one extremal value for
$2 \leq j \leq \lfloor \frac{n}{2} \rfloor$. Moreover, $\frac{\partial
  F}{\partial j}\left(n,\frac{n}{2}\right) <\frac{2}{ \lceil \frac{n}{2} \rceil }-1+
\ln(\frac{2}{nB^2})$ is
thus strictly negative,
as soon as $n\geq 4$.
Now let us define $G(j)=2H(n-j)-2H(j)-\ln(n-j)-\ln(B^2)-1$. 
Using the bounds on the Harmonic numbers, we have that
$$ \frac{2}{2n-2j+\frac{1}{1-\gamma}-2}-\frac{2}{2j+\frac{1}{3}}
< G(j)-\ln\left(\frac{n-j}{j^2}\right)+1+ln(B^2) 
< \frac{2}{2n-2j+\frac{1}{3}}-\frac{2}{2j+\frac{1}{1-\gamma}-2}
$$
Then, on the one hand, we have that
$\frac{2}{2n-2j+\frac{1}{1-\gamma}-2}-\frac{2}{2j+\frac{1}{3}}$
is increasing for $2\leq j \leq 
\frac{n}{2}$ so that its minimal value is 
$M_i(n) = \frac{2}{2n-6+\frac{1}{1-\gamma}}-\frac{6}{13}$ at
$j=2$. Finally, $M_i(n) > -\frac{6}{13}$ if we let $n$ go to the infinity. 

On the other hand, 
$\frac{2}{2n-2j+\frac{1}{3}}-\frac{2}{2j+\frac{1}{1-\gamma}-2}$ is
also increasing and therefore its maximal value is 
$M_s(n) = \frac{2(-4+7\gamma)}{(n-n\gamma-1+2\gamma)(3n+1)}$ at $j=n/2$.
Finally, $M_s(n) \leq \frac{2(7\gamma-4)}{13(3-2\gamma)}$, its value
at $n=4$.

Then, the monotonicity of $G$ and its bounds 
prove that the maximal value of $F(n,j)$ is found for $j^*$
between the solutions $j_i$ and $j_s$ of the two equations below:
\begin{gather}
\ln\left(\frac{n-j_i}{j_i^2}\right) = 1 +
\ln(B^2)+\frac{6}{13}.\\
\ln\left(\frac{n-j_s}{j_s^2}\right) = 1 +
\ln(B^2)-\frac{2(7\gamma-4)}{13(3-2\gamma)}.
\end{gather}
This proves in turn that
$$j* \leq \max\lbrace 0 ; \frac{-1+\sqrt{1+2\delta B^2n}}{\delta B^2}
\rbrace$$
where $\delta = 2 e^{1-\frac{2(7\gamma-4)}{13(3-2\gamma)}} \approx 5.418236$.

Now for the complexity, we use the following recursive scheme to
compute the bound:
\[ \begin{cases}
\log(F(n,0))=\frac{n}{2}\log(nB^2) \\
\log(F(n,j+1))=\log(\frac{F(n,j)}{B}) + \log(\frac{n-j}{j+1})
+\frac{n-j-1}{2}\log(n-j-1)-\frac{n-j}{2}\log(n-j)
\end{cases}
\]
\end{proof}
For instance, if we apply this lemma to matrix \ref{matex} we see that
we just have to look at $F(n,j)$ for $j<\frac{-1+\sqrt{1+2\delta
    B^2n}}{\delta B^2} \approx 1.183$.  

 \section{Eigenvalue bound on the minimal polynomial}\label{sssec:minpeigen}

 For the minimal polynomial the Hadamard bound may also be used, but
 is too pessimistic an estimate, in particular when the degree is
 small. Indeed, one can use Mignotte's bound on the minimal
 polynomial, as a factor of the characteristic polynomial. 
 There, $||\minpoly_A||_\infty \leq 2^d ||C_A||_\infty$, see
 \cite[Theorem 4]{Mignotte:1989:poly}. This
 yields that the bit size of the largest
 coefficient of the
 minimal polynomial is only $d$ bits less than that of the
 characteristic polynomial.

 Therefore, one can rather use a bound on the eigenvalues 
 determined by consideration e.g. of 
 Gershgörin disks and ovals of Cassini (see  
 \cite{Varga:2004:GersC} for more details on the regions containing
 eigenvalues, and \cite[Algorithm OCB]{jgd:2001:JSC} for a blackbox
 algorithm efficiently computing such a bound).
 This gives a bound on the coefficients of the minimal polynomial 
 of the form $\beta^d$
 where $\beta$ is a bound on the eigenvalues 
 and $d$ is the degree of the minimal polynomial. 

 %
 %
 We can then use the following lemma to bound the coefficients
 of the minimal polynomial:

 \begin{lemma} \label{prop:coeffbound}
 Let $A \in \C^{n \times n}$ with its spectral radius bounded by
 $\beta \geq 1$.
 Let $\minpoly_A(X) = \sum_{k=0}^d m_i X^i$. 
 Then 
\[\forall i, \ |m_i| \leq \begin{cases}
\beta^d  & \text{if}~d \leq \beta \\
min \lbrace \sqrt{\beta d}^d~;~\sqrt{\frac{2}{d \pi}}2^d\beta^d \rbrace & \text{otherwise}
\end{cases}\]
\end{lemma} 
This improves the bound given in \cite[Proposition 3.1]{jgd:2001:JSC} 
by a factor of $\log(d)$ when $d >> \beta$.

 \begin{proof} Expanding the minimal polynomial yields
$|m_i| \leq {d \choose i}
   \beta^{d-i}$ by e.g. \cite[Theorem IV.\S 4.1]{Mignotte:1989:poly}.
   Then, if $d \leq \beta$, we bound the latter by $d^i \beta^{d-i}$.

   Now, when $d > \beta$, we get the fist bound in two steps:
   first, for $i \leq \frac{d}{2}$, we bound the binomial factor by
   $d^i$ and thus get ${d \choose i}\beta^{d-i} \leq d^i
   \beta^{\frac{d}{2}-i}\beta^{\frac{d}{2}} <
   d^\frac{d}{2}\beta^\frac{d}{2}$ since $d > \beta$ ; second, for  $i
   > \frac{d}{2}$, we bound the binomial factor by
   $d^{d-i}$ and thus get ${d \choose i}\beta^{d-i} \leq d^{d-i}
   \beta^{d-i} <
   d^\frac{d}{2}\beta^\frac{d}{2}$.
 
   The second bound, when $d \geq \beta$ is obtained by bounding the
   binomial coefficients by the middle one, ${d \choose \frac{d}{2}}$,
   and using  St\u{a}nic\u{a}'s bound \cite{Stanica:2001:binomial} 
   on the latter.
   This gives that ${d \choose i} \beta^{d-i} \leq
   \frac{1}{\sqrt{2\pi}}\sqrt{\frac{4}{d}} 2^{\frac{d}{2}}2^{\frac{d}{2}}\beta^d$.
 \end{proof}

 For matrices of constant size entries, both $\beta$ and $d$ are $\bO(n)$.
 However, when $d$ and/or $\beta$
 is small relative to $n$ (especially $d$) this may be a striking
 improvement over the Hadamard bound since the length of latter
 would be of order $n\log(n)$ rather than $d\log(\beta)$.

 This is the case e.g. for the Homology matrices in the experiments of
 \cite{jgd:2001:JSC}. 
 Indeed, for those, $A A^t$, the Wishart matrix of $A$,
 has very small minimal polynomial degree and has some other useful properties
 which limit $\beta$
 (e.g. the matrix $AA^t$ is diagonally dominant).
 For example, the most difficult computation 
of \cite{jgd:2001:JSC}, is that of the $25605\times 69235$ 
matrix \verb!n4c6.b12! which
has a degree $827$ minimal polynomial with eigenvalues bounded by
$117$. The refinement of lemma \ref{prop:coeffbound}
yields there a gain {\em in size}
on the one of \cite{jgd:2001:JSC} of roughly $5\%$. In this case, this
represents saving $23$ modular projections and an hour of computation. 
\section{Conclusion}
We have presented in this note bounds on the coefficient of the
characteristic and minimal polynomials of a matrix.
Moreover, we give algorithms with low complexity computing even
sharper estimates on the fly.

The refinements given here are only constant with regards to previous
results but yield significant practical speed-ups.

\begin{thebibliography}{10}

\bibitem{jgd:2005:charp}
Jean-Guillaume Dumas, Cl\'ement Pernet, and Zhendong Wan.
\newblock Efficient computation of the characteristic polynomial.
\newblock In Manuel Kauers, editor, {\em Proceedings of the 2005 International
  Symposium on Symbolic and Algebraic Computation, Beijing, China}, pages
  140--147. ACM Press, New York, July 2005.

\bibitem{jgd:2001:JSC}
Jean-Guillaume Dumas, B.~David Saunders, and Gilles Villard.
\newblock On efficient sparse integer matrix {Smith} normal form computations.
\newblock {\em Journal of Symbolic Computation}, 32(1/2):71--99, July--August
  2001.

\bibitem{Gantmacher:1959:TMone}
Feliks~Rudimovich Gantmacher.
\newblock {\em The Theory of Matrices}.
\newblock Chelsea, New York, 1959.

\bibitem{VonzurGathen:1999:MCA}
Joachim~{von zur} Gathen and J{\"u}rgen Gerhard.
\newblock {\em Modern Computer Algebra}.
\newblock Cambridge University Press, New York, NY, USA, 1999.

\bibitem{Giesbrecht:2002:CRF}
Mark Giesbrecht and Arne Storjohann.
\newblock Computing rational forms of integer matrices.
\newblock {\em Journal of Symbolic Computation}, 34(3):157--172, September
  2002.

\bibitem{Mignotte:1989:poly}
Maurice Mignotte.
\newblock {\em Math{\'e}matiques pour le calcul formel}.
\newblock Presses Universitaires Fran{\c c}aises, 1989.

\bibitem{Qi:2005:harmonic}
Feng Qi, Run-Qing Cui, Chao-Ping Chen, and Bai-Ni Guo.
\newblock Some completely monotonic functions involving polygamma functions and
  an application.
\newblock {\em Journal of mathematical analysis and applications},
  310(1):303--308, 2005.

\bibitem{Storjohann:2000:thesis}
Arne Storjohann.
\newblock {\em Algorithms for Matrix Canonical Forms}.
\newblock PhD thesis, Institut f{\"u}r Wissenschaftliches Rechnen, ETH-Zentrum,
  Z{\"u}rich, Switzerland, November 2000.

\bibitem{Stanica:2001:binomial}
Pantelimon St\u{a}nic\u{a}.
\newblock Good lower and upper bounds on binomial coefficients.
\newblock {\em Journal of Inequalities in Pure and Applied Mathematics},
  2(3):Art. 30, 2001.

\bibitem{Varga:2004:GersC}
Richard~S. Varga.
\newblock {\em Ger\v sgorin and his circles}, volume~36 of {\em Springer Series
  in Computational Mathematics}.
\newblock Springer-Verlag, Berlin, 2004.

\end{thebibliography}

\end{document}